\newcommand{\mytex}{}
\newcommand{\stdpackages}{
  \usepackage{amsmath}
  \usepackage{amssymb}
  \usepackage{amsfonts}
  \allowdisplaybreaks
  \usepackage{amsthm}
  \usepackage{eucal}
  \usepackage{graphicx}
  \usepackage{color}
  \usepackage{multicol} 
  \usepackage{fancyhdr}
  \renewcommand{\headrulewidth}{.0pt}\renewcommand{\footrulewidth}{.0pt}\cfoot{}
  \fancyhead[OL]{\it\theauthor---\today}
  \fancyhead[ER]{\leftmark}
  \fancyhead[OR,EL]{\thepage}
  \fancyfoot[EL,OR]{}

  \newcommand{\draft}{\usepackage[light,first]{draftcopy}\draftcopyName{draft}{350}}
  \newcommand{\labels}{\usepackage{\mytex showlabels}}
  \newcommand{\maple}{\usepackage{maple2e}}
  \newcommand{\makeidx}{\usepackage{makeidx}\makeindex}
  \newcommand{\chicago}{\usepackage{\mytex chicago}\bibliographystyle{\mytex chicago}
    \renewcommand{\refname}{References\thispagestyle{empty}\renewcommand{\refname}{}}}
  \newcommand{\numberlines}{
    \usepackage[mathlines,modulo]{\mytex lineno} 
    \newcommand{\BM}{\begin{linenomath}}
    \newcommand{\EM}{\end{linenomath}}
    \linenumbers
    \modulolinenumbers[5]
  }
  \newcommand{\pdflatex}{
    \definecolor{bluecol}{rgb}{0,0,.5}
    \definecolor{greencol}{rgb}{0,.6,0}
    \usepackage[
    pdftex,
    bookmarks,
    bookmarksnumbered,
    colorlinks,
    urlcolor=bluecol,
    citecolor=bluecol,
    linkcolor=bluecol,
    pagecolor=bluecol,
    pdfborder={0 0 0},
    pdfpagemode=None, 
    pdfauthor={Marc Toussaint}
    ]{hyperref}
    \DeclareGraphicsExtensions{.jpg,.pdf}
    \renewcommand{\r}{\varrho}
    \renewcommand{\l}{\lambda}
    \renewcommand{\L}{\Lambda}
    \renewcommand{\s}{\sigma}
    \renewcommand{\O}{\Omega}
    \renewcommand{\SS}{{\cal S}}
    \renewcommand{\boldsymbol}{}
  }
}
\newcommand{\stdtheorems}{
  \theoremstyle{plain}

  \theoremstyle{definition}
  
  \theoremstyle{remark}

}
\newcommand{\stdstyle}[1]{
  \stdpackages
  \stdtheorems
  \renewcommand{\labelenumi}{\textbf{(\roman{enumi})}}
  \renewcommand{\theenumi}{(\roman{enumi})} 
  \newcommand{\itemdot}{\renewcommand{\labelitemi}{\bf $\cdot$}}
  \newcommand{\enumA}{\renewcommand{\labelenumi}{\textbf{\Alph{enumi}}}}
  \newcommand{\blockindent}{3ex}
  \renewcommand{\baselinestretch}{#1}
  \renewcommand{\arraystretch}{1.2}
  \renewcommand{\topfraction}{1}
  \renewcommand{\bottomfraction}{1}
  \renewcommand{\textfraction}{0}
  \columnsep 5ex
  \parindent 3ex
  \parskip 1ex

  \parindent 0pt
  \topsep 4pt plus 1pt minus 2pt
  \partopsep 1pt plus 0.5pt minus 0.5pt
  \itemsep 2pt plus 1pt minus 0.5pt
  \parsep 2pt plus 1pt minus 0.5pt
  \parskip .5pc

  \setcounter{tocdepth}{3}
  \setcounter{secnumdepth}{3}

  \usepackage{\mytex geometry}
  \geometry{a4paper,hdivide={35mm,*,35mm},vdivide={35mm,*,35mm}}

  \renewenvironment{abstract}
    {\vspace*{5ex}\begin{rblock}\hrule\vspace{2ex}{\bf Abstract.~}\small}
    {\vspace{3ex}\hrule\end{rblock}\vspace{5ex}}
  \usepackage{\mytex mt}
}
\newcommand{\cleardefs}{
  \renewcommand{\article}[2]{}
  \renewcommand{\book}[2]{}
  \renewcommand{\draft}{}
  \renewcommand{\labels}{}
  \renewcommand{\maple}{}
  \renewcommand{\makeidx}{}
  \renewcommand{\chicago}{}
  \renewcommand{\pdflatex}{}
  \renewcommand{\header}{}
}

\newcommand{\article}[2]{
  \documentclass[#1pt,twoside,fleqn]{article}
  \stdstyle{#2}
  \macros
  \newcommand{\mytitle}{
    \thispagestyle{empty}
    \mbox{~}
    \begin{list}{}{\leftmargin6ex \rightmargin6ex \topsep0ex \parsep3ex}\item[]
      \begin{center}
        {\LARGE\bf \thetitle \\}

        \vspace{5ex}
        {\large \theauthor}

        {\footnotesize{\sl \address}\\ \email}

        {\footnotesize \today}

        \vspace{1ex}
        {\small \published}
      \end{center}
    \end{list}
    \renewcommand{\mytitle}{\chapter{\thetitle}}
  }
}
\newcommand{\nips}{
  \documentclass{article}
  \usepackage{\mytex nips2003e,times}
  \stdpackages\macros
}
\newcommand{\ijcnn}{
  \documentclass[10pt,twocolumn]{\mytex ijcnn}
  \stdpackages\macros
  \bibliographystyle{abbrv} 
}
\newcommand{\springer}{
  \documentclass{\mytex springer_llncs}
  \renewcommand{\theenumi}{\alph{enumi}}
  \renewcommand{\labelenumi}{(\alph{enumi})}
  \renewcommand{\labelitemi}{$\bullet$}
  \stdpackages\macros
}
\newcommand{\foga}{
  \documentclass{article} 
  \stdpackages\macros
  \usepackage{\mytex foga-02}
  \usepackage{\mytex chicago}
  \bibliographystyle{\mytex foga-chicago}
}
\newcommand{\book}[2]{
  \documentclass[#1pt,twoside,fleqn]{book}
  \newenvironment{abstract}{\begin{rblock}{\bf Abstract.~}\small}{\end{rblock}}
  \stdstyle{#2}
  \newcommand{\mytitle}{
    \thispagestyle{empty}
    \mbox{~}
    \begin{list}{}{\leftmargin4ex \rightmargin4ex \topsep10ex \parsep3ex}\item[]
      \begin{center}
        {\LARGE \thetitle \\}

        \vspace{8ex}
        {\large \theauthor}

        {\footnotesize{\sl \address}\\ \email}

        {\footnotesize \today}

        \vspace{1ex}
        {\small \published}
      \end{center}
    \end{list}
    \renewcommand{\mytitle}{\chapter{\thetitle}}
  }
  \macros
}

\newcommand{\slides}{
  \documentclass[fleqn]{article}
  \stdpackages
  \stdtheorems
  \renewcommand{\baselinestretch}{1}
  \renewcommand{\arraystretch}{1.2}

  \usepackage{\mytex geometry}
  \geometry{
    a4paper,landscape,
    headheight=30mm,
    headsep=0mm,
    footskip=5mm,
    hdivide={10mm,*,10mm},vdivide={30mm,*,8mm}}

  \columnsep 0mm
  \columnseprule 0pt
  \parindent 0ex
  \parskip 0ex
  \setlength{\itemsep}{8ex}
  \renewcommand{\labelitemi}{\rule[.4ex]{.6ex}{.6ex}~}

  \pagestyle{fancy}
  \renewcommand{\headrulewidth}{0pt} 
  \renewcommand{\footrulewidth}{0pt}
  \renewcommand{\labelenumi}{\textbf{\arabic{enumi}.}~~}
  \newcommand{\theauthor}{Marc Toussaint} 
  \rhead{}
  \lhead{}
  \rfoot{\thepage}

  \definecolor{grey}{rgb}{.9,.9,.9}
  \newcommand{\inverted}{
    \definecolor{main}{rgb}{1,1,1}
    \color{main}
    \pagecolor[rgb]{.3,.3,.3}
  }

  \macros

  \newcommand{\mytitle}{\huge\sf}
}

\newcommand{\poster}{
  \documentclass[fleqn]{article}
  \stdpackages
  \renewcommand{\baselinestretch}{1}
  \renewcommand{\arraystretch}{1.8}

  \usepackage{\mytex geometry}
  \geometry{
    paperwidth=1189mm,
    paperheight=841mm, 
    headheight=0mm,
    headsep=0mm,
    footskip=0mm,
    hdivide={5mm,*,5mm},vdivide={5mm,*,5mm}}

  \setlength{\columnsep}{0ex}
  \columnseprule 3pt
  \renewcommand{\labelitemi}{\rule[.4ex]{.6ex}{.6ex}~}

  \pagestyle{fancy}
  \renewcommand{\headrulewidth}{0pt}
  \renewcommand{\footrulewidth}{0pt}
  \renewcommand{\labelenumi}{\textbf{(\roman{enumi})}}
  \newcommand{\theauthor}{Marc Toussaint}
  \rhead{}
  \lhead{}
  \rfoot{}

  \definecolor{grey}{rgb}{.9,.9,.9}
  \newcommand{\inverted}{
    \definecolor{main}{rgb}{1,1,1}
    \color{main}
    \pagecolor[rgb]{.3,.3,.3}
  }

  \macros
}

\author{Marc Toussaint}

\newcommand{\ini}{Institut f\"ur Neuroinformatik, Ruhr-Universit\"at Bochum, Germany}

\newcommand{\email}{mtoussai@inf.ed.ac.uk}

\newcommand{\address}{
  Institute~for~Adaptive~and~Neural~Computation,
  University~of~Edinburgh, 5~Forrest~Hill,
  Edinburgh~EH1~2QL, Scotland,~UK
}

\newcommand{\published}{}

\newlength{\subsecwidth}

\newcommand{\content}[1]{
}

\newcounter{parac}

\newenvironment{rblock}{
\begin{list}{}{\leftmargin\blockindent \rightmargin\blockindent \topsep-\parskip}\item[]}{\end{list}}

\newcommand{\macros}{
  \newcommand{\0}{{\hat 0}}
  \newcommand{\1}{{\hat 1}}
  \newcommand{\2}{{\hat 2}}
  \newcommand{\3}{{\hat 3}}
  \newcommand{\5}{{\hat 5}}

  \renewcommand{\a}{\ensuremath\alpha}
  \renewcommand{\b}{\beta}
  \renewcommand{\c}{\gamma}
  \renewcommand{\d}{\delta}
    \newcommand{\D}{\Delta}
    \newcommand{\e}{\epsilon}
    \newcommand{\g}{\gamma}
    \newcommand{\G}{\Gamma}
  \renewcommand{\l}{\lambda}
  \renewcommand{\L}{\Lambda}
    \newcommand{\m}{\mu}
    \newcommand{\n}{\nu}
    \newcommand{\N}{\nabla}
  \renewcommand{\k}{\kappa}
  \renewcommand{\o}{\omega}
  \renewcommand{\O}{\Omega}
    \newcommand{\p}{\phi}
    \newcommand{\ph}{\varphi}
  \renewcommand{\P}{\Phi}
  \renewcommand{\r}{\varrho}
    \newcommand{\s}{\sigma}
    \newcommand{\Si}{\Sigma}
  \renewcommand{\t}{\theta}
    \newcommand{\T}{\Theta}
  \renewcommand{\v}{\vartheta}
    \newcommand{\x}{\xi}
    \newcommand{\X}{\Xi}
    \newcommand{\Y}{\Upsilon}

  \renewcommand{\AA}{{\cal A}}
    \newcommand{\BB}{{\cal B}}
    \newcommand{\CC}{{\cal C}}
    \newcommand{\EE}{{\cal E}}
    \newcommand{\FF}{{\cal F}}
    \newcommand{\GG}{{\cal G}}
    \newcommand{\HH}{{\cal H}}
    \newcommand{\II}{{\cal I}}
    \newcommand{\KK}{{\cal K}}
    \newcommand{\LL}{{\cal L}}
    \newcommand{\MM}{{\cal M}}
    \newcommand{\NN}{{\cal N}}
    \newcommand{\OO}{{\cal O}}
    \newcommand{\PP}{{\cal P}}
    \newcommand{\QQ}{{\cal Q}}
    \newcommand{\RR}{{\cal R}}
  \renewcommand{\SS}{{\cal S}}
    \newcommand{\TT}{{\cal T}}
    \newcommand{\uu}{{\cal u}}
    \newcommand{\UU}{{\cal U}}
    \newcommand{\XX}{{\cal X}}
    \newcommand{\YY}{{\cal Y}}
    \newcommand{\SOSO}{{\cal SO}}
    \newcommand{\GLGL}{{\cal GL}}

    \newcommand{\Ee}{{\rm E}}

  \newcommand{\NNN}{{\mathbb{N}}}
  \newcommand{\ZZZ}{{\mathbb{Z}}}
  \newcommand{\RRR}{{\mathbb{R}}}
  \newcommand{\CCC}{{\mathbb{C}}}
  \newcommand{\one}{{{\bf 1}}}
  \newcommand{\eee}{\text{e}}

  \renewcommand{\[}{\Big[}
  \renewcommand{\]}{\Big]}
  \renewcommand{\(}{\Big(}
  \renewcommand{\)}{\Big)}
  \renewcommand{\|}{\big|}
  \newcommand{\<}{{\ensuremath\langle}}
  \renewcommand{\>}{{\ensuremath\rangle}}

  \newcommand{\Prob}{{\rm Prob}}
  \newcommand{\Aut}{{\rm Aut}}
  \newcommand{\cor}{{\rm cor}}
  \newcommand{\corr}{{\rm corr}}
  \newcommand{\cov}{{\rm cov}}
  \newcommand{\sd}{{\rm sd}}
  \newcommand{\tr}{{\rm tr}}
  \newcommand{\Tr}{{\rm Tr}}
  \newcommand{\id}{{\rm id}}
  \newcommand{\Gl}{{\rm Gl}}
  \newcommand{\lag}{\mathcal{L}}
  \newcommand{\inn}{\rfloor}
  \newcommand{\lie}{\pounds}
  \newcommand{\longto}{\longrightarrow}
  \newcommand{\speer}{\parbox{0.4ex}{\raisebox{0.8ex}{$\nearrow$}}}
  \renewcommand{\dag}{ {}^\dagger }
  \newcommand{\h}{{}^\star}
  \newcommand{\w}{\wedge}
  \newcommand{\too}{\longrightarrow}
  \newcommand{\To}{\Rightarrow}
  \newcommand{\Too}{\;\Longrightarrow\;}
  \newcommand{\oto}{\leftrightarrow}
  \newcommand{\ow}{\stackrel{\circ}\wedge}
  \newcommand{\feed}{\nonumber \\}
  \newcommand{\comma}{~,\quad}
  \newcommand{\period}{~.\quad}
  \newcommand{\del}{\partial}
  \newcommand{\point}{$\bullet~~$}
  \newcommand{\doubletilde}{
  ~ \raisebox{0.3ex}{$\widetilde {}$} \raisebox{0.6ex}{$\widetilde {}$} \!\!
  }
  \newcommand{\topcirc}{\parbox{0ex}{~\raisebox{2.5ex}{${}^\circ$}}}
  \newcommand{\topdot} {\parbox{0ex}{~\raisebox{2.5ex}{$\cdot$}}}
  \newcommand{\topddot} {\parbox{0ex}{~\raisebox{1.3ex}{$\ddot{~}$}}}
  \newcommand{\sym}{\topcirc}

  \newcommand{\half}{\frac{1}{2}}
  \newcommand{\third}{\frac{1}{3}}
  \newcommand{\fourth}{\frac{1}{4}}

  \newcommand{\ubar}{\underline}

  \renewcommand{\vec}{\boldsymbol}
  \renewcommand{\_}{\underset}
  \renewcommand{\^}{\overset}
  \renewcommand{\*}{\text{\footnotesize\raisebox{-.4ex}{*}{}}}

  \newcommand{\gto}{{\raisebox{.5ex}{${}_\rightarrow$}}}
  \newcommand{\gfrom}{{\raisebox{.5ex}{${}_\leftarrow$}}}
  \newcommand{\gnto}{{\raisebox{.5ex}{${}_\nrightarrow$}}}
  \newcommand{\gnfrom}{{\raisebox{.5ex}{${}_\nleftarrow$}}}

  \newcommand{\RND}{{\SS}}
  \newcommand{\IF}{\text{if }}
  \newcommand{\AND}{\textsc{and }}
  \newcommand{\OR}{\textsc{or }}
  \newcommand{\XOR}{\textsc{xor }}
  \newcommand{\NOT}{\textsc{not }}
}

\newcommand{\argmin}[1]{\underset{~#1}{\rm argmin}\;}

\newcommand{\kld}[2]{D\big(#1:#2\big)}

\newcommand{\matr}[2]{\left(\begin{array}{#1}#2\end{array}\right)}

\newcommand{\url}[1]{\texttt{#1}}

\newcommand{\hide}[1]{$\ll${\sf{\footnotesize #1}}$\gg$\message{^^JHIDE--Warning!^^J}}
\newcommand{\Hide}{\renewcommand{\hide}[1]{\message{^^JHIDE--Warning (hidden)!^^J}}}

\newcommand{\thetitle}{bla}
\newcommand{\header}{
\begin{document}\mytitle\cleardefs}
\newcommand{\contents}{{\tableofcontents}\renewcommand{\contents}{}}
\newcommand{\footer}{\small\bibliography{\mytex bibs}\end{document}}

\article{10}{1}
\chicago
 \Hide

\newcommand{\EM}{}\newcommand{\BM}{}
\newcommand{\pl}{{\protect\rule[.3ex]{1ex}{1ex}}}
\newcommand{\mi}{\ensuremath{\circ}}
\newcommand{\ze}{\ensuremath{\cdot}}
\newcommand{\rb}[1]{\hspace{4pt}\raisebox{-.5ex}{\rotatebox{90}{#1}}\hspace{4pt}}

\title{Notes on information geometry\\ and evolutionary processes}

\header

\begin{abstract}
  In order to analyze and extract different structural properties of
  distributions, one can introduce different coordinate systems over
  the manifold of distributions. In Evolutionary Computation, the
  Walsh bases and the Building Block Bases are often used to describe
  populations, which simplifies the analysis of evolutionary operators
  applying on populations. Quite independent from these approaches,
  information geometry has been developed as a geometric way to
  analyze different order dependencies between random variables (e.g.,
  neural activations or genes).
  
  In these notes I briefly review the essentials of various coordinate
  bases and of information geometry. The goal is to give an overview
  and make the approaches comparable. Besides introducing meaningful
  coordinate bases, information geometry also offers an explicit way
  to distinguish different order interactions and it offers a
  geometric view on the manifold and thereby also on operators that
  apply on the manifold. For instance, uniform crossover can be
  interpreted as an orthogonal projection of a population along an
  $m$-geodesic, monotonously reducing the $\t$-coordinates that
  describe interactions between genes.
\end{abstract}

\section{Introduction}

Evolution can be understood as a process on the space $\L$ of
distributions over the search $\O$. Essentially, a parent population
can be captured as a (finite) distribution $p \in \L$. Mutation and
recombination operators ($\MM \CC$) applied on the parent population
specify a search (offspring) distribution $q \in\L$. And a (stochastic) selection
operator ($\SS^\m\, \FF\, \SS^\n$) maps $q$ to a new parent population
$p'$. In this view, evolution can be understood as a process
\BM\begin{align*}
p
 ~\stackrel{\MM \CC}\longmapsto~ q
 ~\stackrel{\SS^\m \FF \SS^\n}\longmapsto~ p'
 ~\stackrel{\MM \CC}\longmapsto~ q'
 ~\stackrel{\SS^\m \FF \SS^\n}\longmapsto~ p''
 ~\stackrel{\MM \CC}\longmapsto~ \cdots
\end{align*}\EM

We do not need to go into the details of the indicated recombination,
mutation, and selection operators here. Instead, we would like to
emphasize an information theoretic point of view on this process.
Typically, the mapping $p \mapsto q$ (which one could also call search
heuristic) from the parent population to the search distribution adds
entropy whereas selection $q \mapsto p'$ reduces entropy. Another
interesting observable in this process is the \emph{structure} of the
distributions---by which we mean the mutual information present in
these distributions. For instance, one can show that ordinary mutation
and crossover operators (on a direct genetic representation) generally
reduce mutual information, i.e., destroy structural content that might
have been present in $p$ after selection \cite{toussaint:04-ecj}.

The analysis of the structure of distributions is an important topic
in various areas. In evolutionary computation, the Walsh
spectrum is a prominent way to analyze the structure of $p$, often
with the aim to transport it to $q$. The Walsh coefficients may also
be considered as a way of describing epistasis. In complex systems,
certain mutual information measures are often used to define the
structuredness (in their terms: complexity) of dynamics systems
\cite{langton:90,sporns-tononi:02}.

In these notes, I want to briefly review the information geometric way
to describe the structure of a distribution \cite{amari:99,amari:01}
and relate it to the field of evolutionary computation. The first step
is simply to present the coordinates introduced by Walsh coefficients
side-by-side with those used in information geometry to make them
comparable. This gives an intuition about the ``bases'' over which
distributions can be analyzed and reveals, for instance, that the
so-called Building-Block-Basis \cite{chryssomalakos-stephens:04}, as
introduced in Evolutionary Computation, is the same as Amari's
$\eta$-basis. Maybe Amari's $\t$-bases is most interesting in its
capabilities to precisely capture $k$th-order mutual dependencies. It
offers a notion of the ``order-spectrum of mutual information''
alternative to the Walsh spectrum. Eventually, Amari's formalism
allows to completely decompose any distribution into its different
$k$th-oder components.

Finally, the \emph{geometry} introduced over the space of
distributions by Amari gives very insightful interpretations of
distances between distributions. A Pythagoras theorem can be
formulated for the Kullback-Leibler divergence. Under some conditions,
minimizations of the Kullback-Leibler divergence can often be
interpreted as orthogonal projections. This offers a geometric view on
some evolutionary operators.

\section{Notations}

\paragraph{Distributions, $\log$-probabilities, and hypercube bases}

The most direct ``coordinate system'' that can be introduced on the
manifold of distributions is given by the probabilities $p(x)$ for all
$x\in \O$ itself. To preserve notational uniformity with other
coordinate systems we write these numbers as $p_x := p(x)$, which
means that $p_x$ is the $x$-th component of $p \in \L$ in the direct
basis. Because of the normalization constraint $\sum_x p_x = 1$, these
are only $|\O|-1$ independent coordinates.

Clearly, instead of using $p_x$ as coordinates, one can also use their
$\log$'s $l_x := -\log p_x$. Taking the log of probabilities is, very
roughly spoken, related to changing to entropic units. (Note the
definition of the entropy of $p$ as $H(p) = -\sum_x p_x \log p_x = {\rm
  E}_p \{l_x\}$.) Thus, coordinates that have some ``entropic
meaning'' (i.e., are related to information theoretic measures like
entropy, mutual information, or Kullback-Leibler divergence) will be
based on these log quantities. Namely, this will be the $\t$-coordinate
system introduced by Amari (see \citeNP{amari:99,amari:01}).

In the following we will speek of bases of coordinate systems.
Essentially, what we mean are basis functions, similar to the sine
and cosine in the Fourier transform. For illustration, we will always
think of $\O$ as the hypercube; the basis function then correspond to
``colorings'' of the hypercube with function values (mostly $1$, $0$,
or $-1$). E.g., if $e_i:~ \O=\{0,1\}^3 \to \{1,0,-1\}$ is the $i$-th
basis function, then the $i$-th coordinate of a distributions $p$ in
this coordinate system is the convolution of $p$ with $e_i$: $p_i =
\<e_i,p\> := \sum_{x\in\O} e_i(x)\, p(x)$. We illustrate such basis
functions by 3D-hypercubes, \raisebox{-3mm}{\begin{picture}(0,0)%
\includegraphics{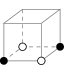}%
\end{picture}%
\setlength{\unitlength}{1184sp}%
\begingroup\makeatletter\ifx\SetFigFont\undefined%
\gdef\SetFigFont#1#2#3#4#5{%
  \reset@font\fontsize{#1}{#2pt}%
  \fontfamily{#3}\fontseries{#4}\fontshape{#5}%
  \selectfont}%
\fi\endgroup%
\begin{picture}(1020,904)(6241,-4928)
\end{picture}%
\setlength{\unitlength}{1pt}
}
where the bullet corresponds to $1$, the circle to $-1$ and empty
vertices to $0$.

The basis of direct coordinate system is the $\d$-basis: the set of
all hypercubes where only one vertex is $1$ and all others are $0$.

\paragraph{Marginals over $k$-tuples of variables and schemata}

In the following, we will also need a compact notation for the
different marginals of a distribution. Let $\O$ be a product space
$\O=\O^1 \times \cdots \times \O^n$ such that we can define the
marginals of a distributions $p$ over single variables but also pairs,
triples, and $k$-tuples of variables. We use indices $i,j,..\in I=
\{1,..,n\}$ to indicate variables and write the marginals as $p^{ij..}$,
\BM\begin{align*}
p^{ij..}(a,b,..)=\Pr\{x_i=a,~ x_j=b,~ ..\} ~.
\end{align*}\EM
The set of all possible marginals is given by considering all single
indices $i$, all pairs $i<j$, all triples $i<j<k$, etc. To simplify
notation (e.g., summation over such objects), we collect all these
tuples of indices in a set
\BM\begin{align*}
A
&= I
  ~\cup~ \{ (i,j) ~|~ i<j \in I \}
  ~\cup~ \{ (i,j,k) ~|~ i<j<k \in I \}
  ~\cup~ \cdots ~\cup~ \{ (1,2,..,n) \} \\
&=\{1,..,n,~ (1,2),(1,3)..,(1,n),(2,3),(2,4)...,(n-1,n),~
    (1,2,3),..,~ (1,2,3,..,n)\} ~.
\end{align*}\EM
In that way, all marginals of $p$ are given as $p^a$ for $a \in
A$. Note that $|A|=|\O|-1$.

Besides using $a \in A$ to indicate a marginal, one can equivalently
use the schemata notation of length-$n$ strings in $\{\*,d\}^n$:
For a given $a$, the corresponding schema is the string of all
$\*$'s except for those positions indicated in the tuple $a$. E.g.,
for $n=6$:
\BM\begin{align*}
p^{245} \equiv p^{\*d\*dd\*}
\end{align*}\EM

\section{Walsh, $\eta$-, $\theta$-, Building Block, and Haar bases}

Table \ref{tabBases} captures the basics of the Walsh, $\eta$-,
$\theta$-, and Haar bases. In all cases, the coordinate system is
defined by the basis functions $e_i$ depicted for the 3D-case as
hypercubes. Actually, these 3D illustrations of the basis functions
$e_i$ are already sufficient to infer the basis functions for all $n$
since they are constructed in a very systematic way---which seems
obvious by simply looking at them and becomes rigorous by considering
the transformation matrices into these coordinates systems:

The transformation matrices map linearly (mod 2) from the direct
coordinates $p_x$ to the new coordinates. E.g., in the Walsh case,
$w_y = \sum_x W_{yx} p_x$. The rows in these matrices correspond to
the basis functions $e_y = W_{y\cdot}$. An important property is that
in all cases (except the Haar bases!), the transformation matrices can
be constructed by repeated tensor products of a 2D matrix. For
instance, for $n=2$ in the Walsh case:
\BM\begin{align*}
W^{n=2}
 = \matr{rrrr}{1&1&1&1\\1&-1&1&-1\\1&1&-1&-1\\1&-1&-1&1}
 = \matr{rr}{1&1\\1&-1} \otimes \matr{rr}{1&1\\1&-1}
 =: \matr{rr}{1&1\\1&-1}^{\!\!\otimes 2}
\end{align*}\EM
Here, we introduced the superscript notation ${}^{\otimes n}$ to
indicate the $n$-fold tensor product.

\newcommand{\inclfig}[1]{\begin{minipage}[t]{35mm}\raisebox{-43mm}{\input{figtexs/#1}}\end{minipage}}

\begin{table}
\begin{tabular}{@{}p{43mm}rr@{}}
{\bf Walsh} \newline
 $w_y = \sum_x W_{yx}\, p_x$\newline
 $p_x = \frac{1}{n}\, \sum_y W_{xy} w_y$\newline
 $W_{yx} = (-1)^{|x\, \AND y|}$\newline
 $~~~~  = \matr{cc}{\pl & \pl \\ \pl & \mi}^{\!\!\otimes n}$\newline
 $W^{-1} = \frac{1}{n}\, W$
 & \begin{tabular}[t]{@{}c@{\,}|@{}c@{}c@{}c@{}c@{}c@{}c@{}c@{}c@{}}
    & \rb{000} & \rb{001} & \rb{010} & \rb{011} & \rb{100} & \rb{101} & \rb{110} & \rb{111} \\
\hline
000 & \pl & \pl & \pl & \pl & \pl & \pl & \pl & \pl \\
001 & \pl & \mi & \pl & \mi & \pl & \mi & \pl & \mi \\
010 & \pl & \pl & \mi & \mi & \pl & \pl & \mi & \mi \\
011 & \pl & \mi & \mi & \pl & \pl & \mi & \mi & \pl \\
100 & \pl & \pl & \pl & \pl & \mi & \mi & \mi & \mi \\
101 & \pl & \mi & \pl & \mi & \mi & \pl & \mi & \pl \\
110 & \pl & \pl & \mi & \mi & \mi & \mi & \pl & \pl \\
111 & \pl & \mi & \mi & \pl & \mi & \pl & \pl & \mi \\
\end{tabular}
  &\begin{minipage}[t]{35mm}\raisebox{-43mm}{\input{figtexs/walsh}}\end{minipage} \\\hline
{\bf Amari's $\eta$ / BBB} \newline
 $\eta_a = \sum_x \bar B_{ax} p_x$\newline
 $~~~~ = \sum_x (B^{-1})^T_{ax} p_x$ \newline
 $p_x = \sum_a B^T_{xa} \eta_a$\newline
 $\bar B = (B^{-1})^T = \matr{cc}{\pl & \pl \\ \ze & \pl}^{\!\!\otimes n}$ \newline
 $\bar B^{-1} = \matr{cc}{\pl & \mi \\ \ze & \pl}^{\!\!\otimes n}$
 & \begin{tabular}[t]{@{}c@{\,}|@{\,}c@{\,}|@{}c@{}c@{}c@{}c@{}c@{}c@{}c@{}c@{}}
  &   & \rb{000} & \rb{001} & \rb{010} & \rb{011} & \rb{100} & \rb{101} & \rb{110} & \rb{111} \\
\hline
$\cdot$ & $\*\*\*$ & \pl & \pl & \pl & \pl & \pl & \pl & \pl & \pl \\
3 & $\*\*1$  & \ze & \pl & \ze & \pl & \ze & \pl & \ze & \pl \\
2  & $\*1\*$ & \ze & \ze & \pl & \pl & \ze & \ze & \pl & \pl \\
23 & $\*11$ & \ze & \ze & \ze & \pl & \ze & \ze & \ze & \pl \\
1  & $1\*\*$ & \ze & \ze & \ze & \ze & \pl & \pl & \pl & \pl \\
13 & $1\*1$ & \ze & \ze & \ze & \ze & \ze & \pl & \ze & \pl \\
12 & $11\*$ & \ze & \ze & \ze & \ze & \ze & \ze & \pl & \pl \\
123 & $111$ & \ze & \ze & \ze & \ze & \ze & \ze & \ze & \pl \\
\end{tabular}
 &\begin{minipage}[t]{35mm}\raisebox{-43mm}{\input{figtexs/eta}}\end{minipage} \\\hline
{\bf Amari's $\t$} \newline
 $\t_a = \sum_x B_{ax} l_x$ \newline
 $l_x = \sum_a \bar B^T_{xa} \t_a$ \newline
 $B =  (\bar B^{-1})^T = \matr{cc}{\pl & \ze \\ \mi & \pl}^{\!\!\otimes n}$ \newline
 $B^{-1} = \matr{cc}{\pl & \ze \\ \pl & \pl}^{\!\!\otimes n}$
 & \begin{tabular}[t]{@{}c@{\,}|@{}c@{}c@{}c@{}c@{}c@{}c@{}c@{}c@{}}
    & \rb{000} & \rb{001} & \rb{010} & \rb{011} & \rb{100} & \rb{101} & \rb{110} & \rb{111} \\
\hline
$\cdot$ & \pl & \ze & \ze & \ze & \ze & \ze & \ze & \ze \\
3   & \mi & \pl & \ze & \ze & \ze & \ze & \ze & \ze \\
2   & \mi & \ze & \pl & \ze & \ze & \ze & \ze & \ze \\
23  & \pl & \mi & \mi & \pl & \ze & \ze & \ze & \ze \\
1   & \mi & \ze & \ze & \ze & \pl & \ze & \ze & \ze \\
13  & \pl & \mi & \ze & \ze & \mi & \pl & \ze & \ze \\
12  & \pl & \ze & \mi & \ze & \mi & \ze & \pl & \ze \\
123 & \mi & \pl & \pl & \mi & \pl & \mi & \mi & \pl \\
\end{tabular}
 &\begin{minipage}[t]{35mm}\raisebox{-43mm}{\input{figtexs/theta}}\end{minipage} \\\hline
{\bf Haar}\newline
please see \cite{khuri:94}
 & \begin{tabular}[t]{@{}c@{\,}|@{}c@{}c@{}c@{}c@{}c@{}c@{}c@{}c@{}}
    & \rb{000} & \rb{001} & \rb{010} & \rb{011} & \rb{100} & \rb{101} & \rb{110} & \rb{111} \\
\hline
000 & \pl & \pl & \pl & \pl & \pl & \pl & \pl & \pl \\
001 & \pl & \pl & \pl & \pl & \mi & \mi & \mi & \mi \\
010 & \pl & \pl & \mi & \mi & \ze & \ze & \ze & \ze \\
011 & \ze & \ze & \ze & \ze & \pl & \pl & \mi & \mi \\
100 & \pl & \mi & \ze & \ze & \ze & \ze & \ze & \ze \\
101 & \ze & \ze & \pl & \mi & \ze & \ze & \ze & \ze \\
110 & \ze & \ze & \ze & \ze & \pl & \mi & \ze & \ze \\
111 & \ze & \ze & \ze & \ze & \ze & \ze & \pl & \mi \\
\end{tabular}
 &\begin{minipage}[t]{35mm}\raisebox{-43mm}{\begin{picture}(0,0)%
\includegraphics{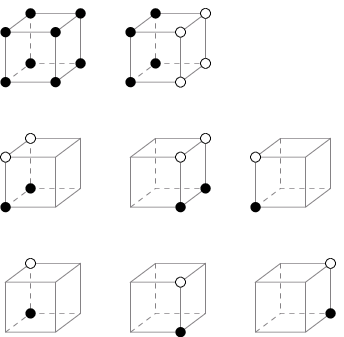}%
\end{picture}%
\setlength{\unitlength}{1579sp}%
\begingroup\makeatletter\ifx\SetFigFont\undefined%
\gdef\SetFigFont#1#2#3#4#5{%
  \reset@font\fontsize{#1}{#2pt}%
  \fontfamily{#3}\fontseries{#4}\fontshape{#5}%
  \selectfont}%
\fi\endgroup%
\begin{picture}(4036,3951)(3233,-4920)
\end{picture}%
\setlength{\unitlength}{1pt}
}\end{minipage}
\end{tabular}
\caption{\label{tabBases}
Overview over the different bases for the space of distributions. The
first column gives the definitions of the transformations and their
inverse. Note that the $\t$-bases is defined in log-space. The 
transformation matrices are illustrated in the section column
for $n=3$ using the symbols $\pl =1$, $\mi=-1$, and $\ze=0$. The third
column illustrates the bases functions $e_y$ (or $e_a$) as colorings
of the hypercube $\{0,1\}^3$. Note that the basis functions
correspond to rows of the transformation matrix. The 1-norm $|x\, \AND y|$ 
of the \AND of two binary strings counts the 1-bits that they have in common.
}
\end{table}

Table \ref{tabBases} summarizes the most important properties of these
transformation matrices: their closed form expression, their tensor
product construction, and their inverse. When looking at the table one
should first observe the self-similar regularity of the transformation
matrices, which stems from their definition of repeated tensor
products. The meaning of the various bases become more intuitive when
looking at the hypercube illustrations of the basis. The Walsh bases,
e.g., can nicely be compared to a Fourier basis: $e_{000}$ corresponds
to the constant function $1$, $e_{001},e_{010},e_{100}$ could be view
as sinus functions along the $x$-, $y$-, and $z$-axes, respectively;
$e_{011},e_{101},e_{110}$ are products of sinus functions---and
capture 2nd order dependencies; and $e_{111}$ is the ``highest
frequency'' bases function capturing 3rd order dependencies.

The $\eta$-bases captures certain marginals relative to the all-1s string:
\BM\begin{align*}
\eta_a=p^a(11..) ~.
\end{align*}\EM
These can be thought of the marginals over all possible
Building-Blocks---thus it is also called the Building-Block-Bases
(BBB, cf.\ \citeNP{chryssomalakos-stephens:04}). This marginalization becomes apparent
in the hypercube colorings as the abundance of zeros (non-colored
vertices and dots in the matrix).

The $\t$-bases combines the ``frequency'' idea of the Walsh
bases with the marginalization: The highest order bases function
$e_{123}$ is analogous to the Walsh bases $e_{111}$ and detects
highest order dependencies. Lower order dependencies though are only
detected on a marginal.

However, note that the $\t$ bases is defined in log-space, $\t_a =
\sum_x B_{ax} \log p_x$. We will find some implications of this in the next
section. Note that the transformation matrices of the $\eta$-
(Building-Block-) and the $\t$-bases are related via $B =  (\bar
B^{-1})^T$.

For completeness, we also indicated the Haar bases in table
\ref{tabBases}. It can not be derived as repeated tensor products and
we do not discuss it any further here. One argument made about the
Haar bases \cite{khuri:94} is that the transformation matrix incorporates
a lot of 0s. Thus, the coefficients are more efficient to compute as
the Walsh coefficients. We add here that the ratio of zeros in the
$\eta$ and $\t$ transformation matrices is $1-(3/4)^{n-1}$ and
approaches $1$ exponentially with the dimension $n$.

\section{Mathematical structure on the manifold $\L$}

In this section we want to develop a more geometric view on the
manifold of distributions, following \cite{amari:99,amari:01}. This
geometry will put a special emphasis on the $\eta$- and $\t$-bases.

\paragraph{$m$- and $e$-geodesics}

An essential ingredient to describe the geometry of a manifold is the
definition of the notion of ``straight lines'', or geodesics, connecting two
points in the manifold. In the case of the manifold of distributions,
there exist at least two ways of defining a straight path connecting two
distributions $q$ and $r$: the one being the linear mixture in direct
coordinates $p_x$, the other being the linear mixture in $\log$
coordinates $l_x$,
\BM\begin{align*}
\text{$m$-geodesic:}\qquad& p(x) = (1\!-\!\a)\, q(x) + \a\, r(x) ~,\\
\text{$e$-geodesic:}\qquad& \log p(x) = (1\!-\!\a)\, \log q(x) + \a\, \log r(x) - \psi(x) ~.
\end{align*}\EM
Here $m$ means \emph{mixture} and $e$ means \emph{exponential}. The
additional term $\psi(x)$ in the $e$-geodesic is necessary to preserve
the normalization of $p(x)$.

The fact that there exist two ways of defining geodesics means that
there exist two meaningful \emph{affine connections} on the manifold.
Both define a notion of
flatness: we say that a $m$-geodesic is $m$-flat and a $e$-geodesic
is $e$-flat.

It turns out that the coordinate lines (and planes, hyperplanes, etc.)
of $\eta$ are $m$-flat and those of $\t$ are $e$-flat. The former is
obvious, since an $m$-geodesic can equivalently be written in the
$\eta$ coordinate system as $\eta_a(p) = (1\!-\!\a)\, \eta_a(q) + \a\,
\eta_a(r)$. The second becomes apparent when realizing that the Taylor
expansion of $\log p$ reads
\BM\begin{align*}
l_x
 = \sum_i \t_i x_i
 + \sum_{i<j} \t_{ij}\, x_i x_j
 + \sum_{i<j<k} \t_{ijk}\, x_i x_j x_k
 + \cdots + \t_{1..n}\, x_1..x_n - \psi
 = \sum_{a\in A} \t_a X^a - \psi
\end{align*}\EM
where $X^a$ is the product of the components $x_{i_1} x_{i_2}\cdots x_{i_k}
\in \{0,1\}$ when $a=(i_1,i_2,..,i_k)$. Thus, an $e$-geodesic is
written, in the $\t$ coordinate system, simply as $\t_a(p) = (1\!-\!\a)\,
\t_a(q) + \a\, \t_a(r)$.

\paragraph{Fisher metric, Kullback-Leibler divergence}

On this manifold $\L$, there is a metric defined, the \emph{Fisher
  metric}.  In \emph{arbitrary} coordinates $v_i$ (it could be any of
the Walsh, log, $\eta$-, or $\t$-coordinates), it reads
\BM\begin{align*}
g_{ij}(p) = {\rm E}\left\{ \frac{\del \log p}{\del v_i}\, \frac{\del \log p}{\del v_j}\right\} ~.
\end{align*}\EM
Some intuition can be gained by realizing that, locally, the distance
measured by the Fisher metric coincides with the distance measured by
the Kullback-Leibler divergence:\footnote{ The Kullback-Leibler
  divergence $\kld{p}{q}$ (also called relative entropy or divergence)
  is a measure for the loss of information (or gain of entropy) when a
  \emph{true} distribution $p$ is approximated by a model
  distributions $q$. For example, when $p(x,y)$ is approximated by
  $p(x)\,p(y)$ one looses information on the mutual dependence between
  $x$ and $y$.  Accordingly, the relative entropy
  $\kld{p(x,y)}{p(x)\,p(y)}$ is equal to the mutual information
  between $x$ and $y$. Generally, when \emph{knowing} the real
  distribution $p$ one needs on average $H(p)$ (entropy of $p$) bits to
  describe a random sample. If, however, we know only an approximate
  model $q$ we would need on average $H(p) + \kld{p}{q}$ bits to
  describe a random sample of $p$.  The loss of knowledge about the
  true distribution induces an increase of entropy and thereby an
  increase of average description length for random samples. }
Consider a point $p \in \L$ and a nearby point $p+\d p$. When we
measure the squared length $\<\d p,\d p\>$ of the variation $\d p$ by
the Kullback-Leibler divergence we find \BM\begin{align*} \<\d p,\d
  p\> = \kld{p}{p+\d p} = {\rm E}\left\{ \log p - \log (p+\d
    p)\right\} \ddot=~ {\rm E}\left\{- \frac{\d p}{p} + \frac{\d
      p^2}{p^2} \right\} = {\rm E}\left\{\frac{\d p^2}{p^2} \right\}
  ~.
\end{align*}\EM
Here, the 2nd-order approximation stems from the Taylor expansion of
$\log(p+\d p)$ and ${\rm E}\{\d p/p\} =0$ since $\sum_x \d p(x)=0$ to
preserve normalization. Note that, in this infinitesimal neighborhood,
the Kullback-Leibler divergence becomes symmetric. Generalizing this
to two small variations $\d_1 p= \del_{v_i} p := \frac{\del p}{\del v_i}$ and $\d_2
p= \del_{v_j} p := \frac{\del p}{\del v_j}$ induced by small shifts along some
coordinates we have
\BM\begin{align*}
\<\del_{v_i} p,\del_{v_j} p\>
 =  {\rm E}\left\{\frac{\del_{v_i} p}{p}\, \frac{\del_{v_i} p}{p} \right\}
 =  {\rm E}\left\{\frac{\del \log p}{\del v_i}\, \frac{\del \log p}{\del v_i} \right\}
\end{align*}\EM
and retrieve the Fisher metric. In turn, the Fisher metric can also be derived by considering the second order derivatives of the Kullback-Leibler divergence:
\BM\begin{align*}
g_{ij}(q) = \frac{1}{2}\, \frac{\del}{\del v_i}\frac{\del}{\del v_j} \kld{p}{p+\d v}\Big|_{\d v=0} ~.
\end{align*}\EM

\paragraph{Orthogonality of $\eta$ and $\t$, the Pythagoras}

\begin{figure}
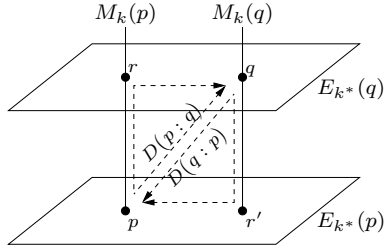
\center
\input figtexs/pythagoras
\caption{\label{figPy}
The Pythagoras in the case when a certain $k$-cut is used to define
the $m$- and $e$ geodesics connecting to $r$, respectively $r'$. It
holds: $\kld{p}{q} = \kld{p}{r} + \kld{r}{q}$ and $\kld{q}{p} = \kld{q}{r'} + \kld{r'}{p}$.
}
\end{figure}

The coordinate systems $\eta$ and $\t$ have a crucial property w.r.t.\ 
the Fisher metric---they are mutually orthogonal: At any point $p$ in
the manifold the variations induced by shifts along $\t$ and $\eta$
coordinates fulfill
\BM\begin{align*}
\<\del_{\t_a} p, \del_{\eta_b} p\> = \d_{ab} ~,
\end{align*}\EM
where $\d_{ab}$ is the Kronecker delta. Based on this one can derive a
Pythagoras theorem: Let $p$, $r$ and $q$ be three distributions where
the $m$-geodesic connecting $p$ and $r$ is orthogonal to the
$e$-geodesic connecting $r$ and $q$, then
\BM\begin{align*}
\kld{p}{q} = \kld{p}{r} + \kld{r}{q} ~.
\end{align*}\EM
Please figure \ref{figPy} for an illustration.

\paragraph{$k$-cuts}

Let $k$ denote an order of interactions that we are interested in.
Then, the coordinates split into those describing interactions of
order $\le k$ and those describing interactions of order $> k$,
\BM\begin{align*}
\vec \eta_k &:= (\text{all $\eta_a$ of order $|a| \le k$}) ~,\\
\vec \t_{k^*} &:= (\text{all $\t_a$ of order $|a| >k$}) ~.
\end{align*}\EM

These can be mixed into a new coordinate system $(\vec \eta_k,\vec
\t_{k^*})$. The point is that those dimensions spanned by $\vec
\eta_k$ are orthogonal to those spanned by $\vec \t_{k^*}$. To
simplify the discussion we call $\vec \eta_k$ \emph{marginals}
(although they include marginals over $k$-tuples of variables) and
$\vec\t_{k^*}$ \emph{higher order interactions}. Keeping the marginals
$\vec \eta_k$ constant defines $m$-flat sub-manifolds $M_k(\eta_k)$,
which are disjoint for different $\vec \eta_k$ and cover all $\L$.
Keeping higher order interactions $\vec \t_{k^*}$ constant defines
$e$-flat sub-manifolds $E_{k^*}(\t_{k^*})$, which are disjoint for
different $\vec \t_{k^*}$ and cover all $\L$.

\paragraph{Complete decomposition of different order interactions}

\begin{figure}\center
\input{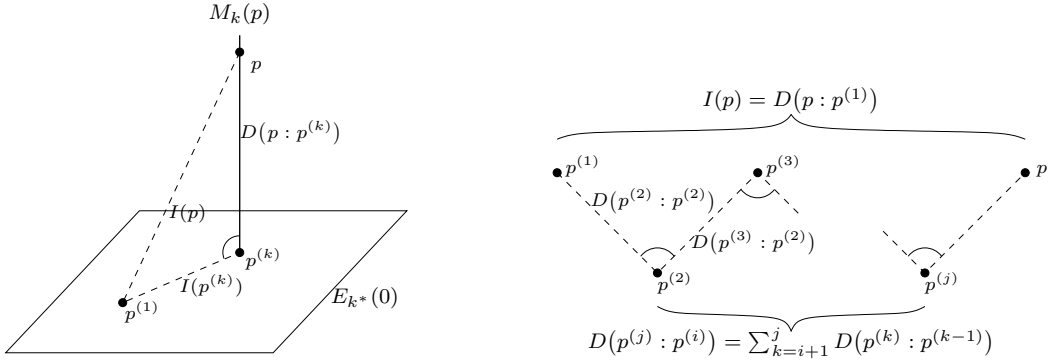}\hfill
\input figtexs/cuts
\caption{\label{figDecomp}
  The left figure illustrates a distribution $p$ and its $k$th-order
  reduction $p^{(k)}$: It is the orthogonal projection of $p$ along
  $M_k(p)$ onto $E_{k^*}(0)$. The ``distance'' $D(p:p^{(k)})$ measures
  ``norm'' of $\vec\t_{k^*}$, i.e., it measures the amount of mutual
  information of order higher than $k$. The right figure illustrates
  the complete decomposition of $p$ in reductions $p^{(k)}$ of all
  orders. Every projection from $p^{(k)}$ to $p^{(k-1)}$ is an
  orthogonal projection onto $E_{(k-1)^*}(0)$. Every ``distance''
  $D(p^{(k)}:p^{(k-1)})$ measures the mutual information specifically
  of order $k$. }
\end{figure}

Given a distribution $p$, we define its $k$th order reduction
$p^{(k)}$ as the distribution with same marginals $\vec \eta_k(p)$ as
$p$ but vanishing higher order interactions $\vec \t_{k^*}=0$,
\BM\begin{align*}
  p^{(k)} = (\vec \eta_k(p),\vec \t_{k^*}=0) ~.
\end{align*}\EM
That is, $p^{(k)}$ is the same distributions as $p$ except that all
interactions of order $>k$ have been canceled. We call $p^{(k)}$ the
$k$th-order reduction of $p$. Given the Pythagoras it should be clear
that $p^{(k)}$ can also be defined as the orthogonal projection of $p$
onto the submanifold $E_{k^*}(0)$ or as the orthogonal projection of
the uniform distribution $p^{(0)}$ onto $M_k(\vec\eta_k(p))$, please
see figure \ref{figDecomp} left,
\BM\begin{align*}
p^{(k)}
 = \argmin{q \in E_{k^*}(0)} \kld{p}{q}
 = \argmin{q \in M_k(\vec\eta_k(p))} \kld{q}{p^{(0)}} ~.
\end{align*}\EM
Further, define $D_k(p) = \kld{p^{(k)}}{p^{(k-1)}}$. Then the
Pythagoras allows to decompose the mutual information $I(p)$ in $p$
(i.e., the measure of all interactions in $p$) into a sum of different
order interactions:
\BM\begin{align*}
I(p) = \kld{p}{p^{(1)}} = \sum_{k=2}^n D_k(p)
\end{align*}\EM
Please see figure \ref{figDecomp} right for an illustration.

This result should be highlighted. The given formalism allows to
explicitly distinguish different order interactions between variables
in a distribution and directly assigns coordinates $\t$ to those
different order interactions. The quantities $D_k(p) =
\kld{p^{(k)}}{p^{(k-1)}}$ measure precisely and only the $k$th-order
interactions in entropic units.

For instance, consider three random variables $X_1,\, X_2,\, X_3$
which are pair-wise dependent in the sense $I(X_i|X_j) \not=0$. The
question is whether there exist ``true'' 3rd-order interactions or only
concatenated 2nd-order interactions---in other terms, can they be
described by a Markov process $X_1 \to X_2 \to X_3$. The formalism
gives an answer: if $D_3(p)=0$ it is a Markov process, otherwise there
exist 3rd-order interactions.

\section{Geometric view on evolution operators}

\paragraph{Crossover}

In Evolutionary Algorithms, crossover is one means of mixing a parent
population to an offspring population. Populations can be formalized
as distributions $p$ and a definition of a simple form of crossover
(uniform crossover parameterized with $c\in\RRR$) reads
\BM\begin{align*}
\CC p &= (1-c)\, p + c\, p^{(1)} ~.
\end{align*}\EM
See, for instance, \cite{toussaint:03-gecco-cross} for a general
definition of a crossover operator in more conventional notation and
details of when it reduces to this simple form.

\begin{figure}\center
\input{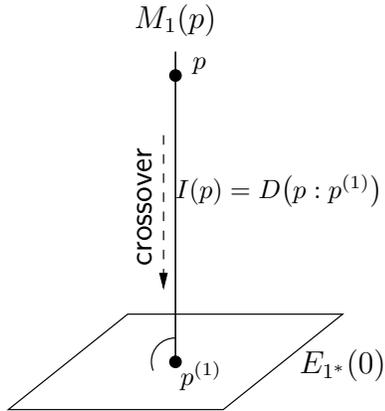}
\caption{\label{figCross}
  Crossover is an operator that takes a step along the projection of
  $p$ towards the first order reduction $p^{(1)}$.}
\end{figure}

This crossover simply mixes the original distribution (or population)
$p$ with its $1$st-order reduction. The $1$st-order reduction is the
product of all single variable marginals, i.e., it is the distribution
with the same marginals (gene frequencies) as $p$ but all dependencies
(gene linkages) between the variables eliminated. From the geometrical
point of view, crossover makes a step along the $m$-geodesic
connecting $p$ and $p^{(1)}$. It can be illustrated as a step along
the projection onto the submanifold $E_{1^*}(p)$, please see figure
\ref{figCross}.

From this view it becomes clear that a reasonable coordinate system to
describe crossover is $(\vec\eta_1,\vec\t_{1^*})$. Crossover does not
change $\vec\eta_1$ (it operates orthogonally to $\eta_1$) but
continuously reduces the $\vec\t_{1^*}$ variables. That $\vec\t_{1^*}$
are reduced and not increased is intuitive from figure \ref{figCross}
(recall that $\t$'s are always positive) and becomes apparent from that
the ``distance'' from $p$ to $p^{(1)}$, $I(p)=\kld{p}{p^{(1)}}$, is a
norm of $\vec\t_{1^*}$.  \hide{ More explicitly, consider the
  derivative of the $\t$-coordinates along the path, i.e., with
  varying $c$, \BM\begin{align*}
    \frac{\del}{\del c} p_x(c) &= p_x^{(1)} - p_x \\
    \frac{\del}{\del c} \t_a(c) &= \sum_x B_{ax}\, \frac{\del}{\del c}
    l_x(c) = \sum_x B_{ax}\, \frac{p_x^{(1)}-p_x}{p_x(c)}
\end{align*}\EM

-- In the $\O=\{0,1\}^2$ two gene case we have
\BM\begin{align*}
& |a|\le 1 \To \eta_a(c) = \eta_a \\
&\eta_{12}(c) = (1-c)\, \eta_{12} + c\, \eta_1\, \eta_2
\end{align*}\EM
with crossover probability $c$. Further
\BM\begin{align*}
\dot\eta_{12}
 &= -\eta_{12} + \eta_1\, \eta_2 \\
\t_{12}
 &= \sum_x B_{(12)x} \log \sum_a B^T_{xa} \eta_a \\
\dot \t_{12}
 &= \sum_x B_{(12)x} \frac{1}{p_x}[\sum_a B^T_{xa} \dot\eta_a]
  = \sum_x B_{(12)x} \frac{B^T_{x(12)} \dot\eta_{12}}{p_x}
  = \dot\eta_{12} \sum_x \frac{B_{(12)x} B^T_{x(12)} }{p_x}
  = \dot\eta_{12} \sum_x \frac{1}{p_x}
\end{align*}\EM
where the $\dot{}$ means the derivative $\del_c$ at $c=0$. Since, on
the path $c\in [0,1]$, $\dot\eta_{12}$ is constant and does not change
the sign, $\t_{12}$ monotonously approaches zero.
}

\paragraph{Max Entropy}

\citeN{wright-et-al:04} recently proposed an evolutionary search
scheme that constructs the new search distribution (offspring
population) via a maximum entropy principle: From the parent
population all second order scheme frequencies are calculated. Then,
from all the distributions which have the same second order schema
frequencies, the new offspring distribution is the one with maximum
entropy.

In our formalism, constraining the schema frequencies corresponds to
fixing $\vec \eta_2$, i.e., constraining the offspring distribution to
the submanifold $M_2(\vec \eta_2)$. The distribution with maximal
entropy in $M_2(\vec \eta_2)$ must have minimal higher order
(3rd-order or higher) interactions $\vec\t_{2^*}$ since interactions
(mutual information) reduce entropy. Thus, the max entropy rule simply
amounts to setting $\vec\t_{2^*}=0$, i.e., choosing
$p^{(2)}=(\vec\eta_2,0)$ as the new offspring distribution.

Again, this can be viewed geometrically as the orthogonal projection
of the parent population $p$ onto $E_{2^*}(0)$ according to
\BM\begin{align*}
\argmin{q \in E_{2^*}(0)} \kld{p}{q}
\end{align*}\EM
or as the orthogonal projection of the uniform distribution $p^{(0)}$
onto $M_2(\vec\eta_2)$
\BM\begin{align*}
\argmin{q \in M_2(\vec\eta_2)} \kld{q}{p^{(0)}} ~.
\end{align*}\EM
This latter way of writing the max entropy principle is quite
intuitive: find the distribution that fulfills the required constraints
(lies on $M_2(\vec\eta_2)$) but is closest to the uniform distribution
$p^{(0)}$.

Eventually, note the strong analogy of the maximum entropy principle
proposed by \cite{wright-et-al:04} and the simple crossover operator
given before: Crossover moves $p$ toward $p^{(1)}$, while the search
heuristic considered by Wright et.\ al.\ chooses $p^{(2)}$ as the new
search distribution.

\section{Discussion}

The methods information geometry provides to analyze and describe the
structure of distributions are deeply grounded in information
theory. For instance, it seems very beneficial to have coordinate
systems for distributions which capture precisely arbitrary $k$th
order interactions between variables and have a direct link to
measures like mutual information and the Kullback-Leibler
divergence. Also the geometric aspects, e.g., that some operations
can be described as orthogonal to certain submanifolds, add to a more
comprehensive picture of the space of distributions. In that sense,
information geometric methods enhance more common approaches in
Evolutionary Computation, like the Walsh bases, in describing the
structure of distributions and operators.

However, the question remains how and whether these methods can be
used to (1) actually propose new heuristic search algorithms or (2) to
provide new theoretical tools to analyze the dynamics of evolutionary
processes.

\subsection*{Acknowledgment}

I would like to thank the German Research Foundation (DFG) for their
funding of the Emmy Noether fellowship TO 409/1-1.

\footer\small
\bibliography{/cygdrive/c/home/tex/bibs}
\end{document}